\documentclass[journal=jacsat,manuscript=article]{achemso}

\usepackage[version=3]{mhchem} 
\usepackage{tabularx}
\usepackage{array} 
\usepackage{booktabs}
\usepackage{comment}
\usepackage{xurl}
\usepackage{etoc}
\usepackage[table]{xcolor}
\usepackage[normalem]{ulem}

\author{Vaibhav Khanna}
\affiliation[UMich]{Department of Chemistry, University of Michigan, Ann Arbor, Michigan 48109, United States}
\author{Bikash Kanungo}
\affiliation[UMich-Mech]{Department of Mechanical Engineering, University of Michigan, Ann Arbor, Michigan 48109, United States}
\author{Vikram Gavini}
\altaffiliation{Department of Materials Science \& Engineering, University of Michigan, Ann Arbor, Michigan 48109, United States}
\affiliation[UMich-Mech]{Department of Mechanical Engineering, University of Michigan, Ann Arbor, Michigan 48109, United States}
\author{Ambuj Tewari}
\affiliation[UMich-Stats]{Department of Statistics, University of Michigan, Ann Arbor, Michigan 48109, United States}
\author{Paul M. Zimmerman}
\affiliation[UMich]{Department of Chemistry, University of Michigan, Ann Arbor, Michigan 48109, United States}
\email{paulzim@umich.edu}

\title[An \textsf{achemso} demo]
  {Examining the Impact of Local Condition Violations on Energy Computations in DFT}

\abbreviations{DFT}

\begin{document}

\begin{center}
\textbf{TOC Graphic}\par
\end{center}
\vspace{5mm}
\begin{figure}[!h]
    \begin{center}
    \includegraphics[height=4.5cm]{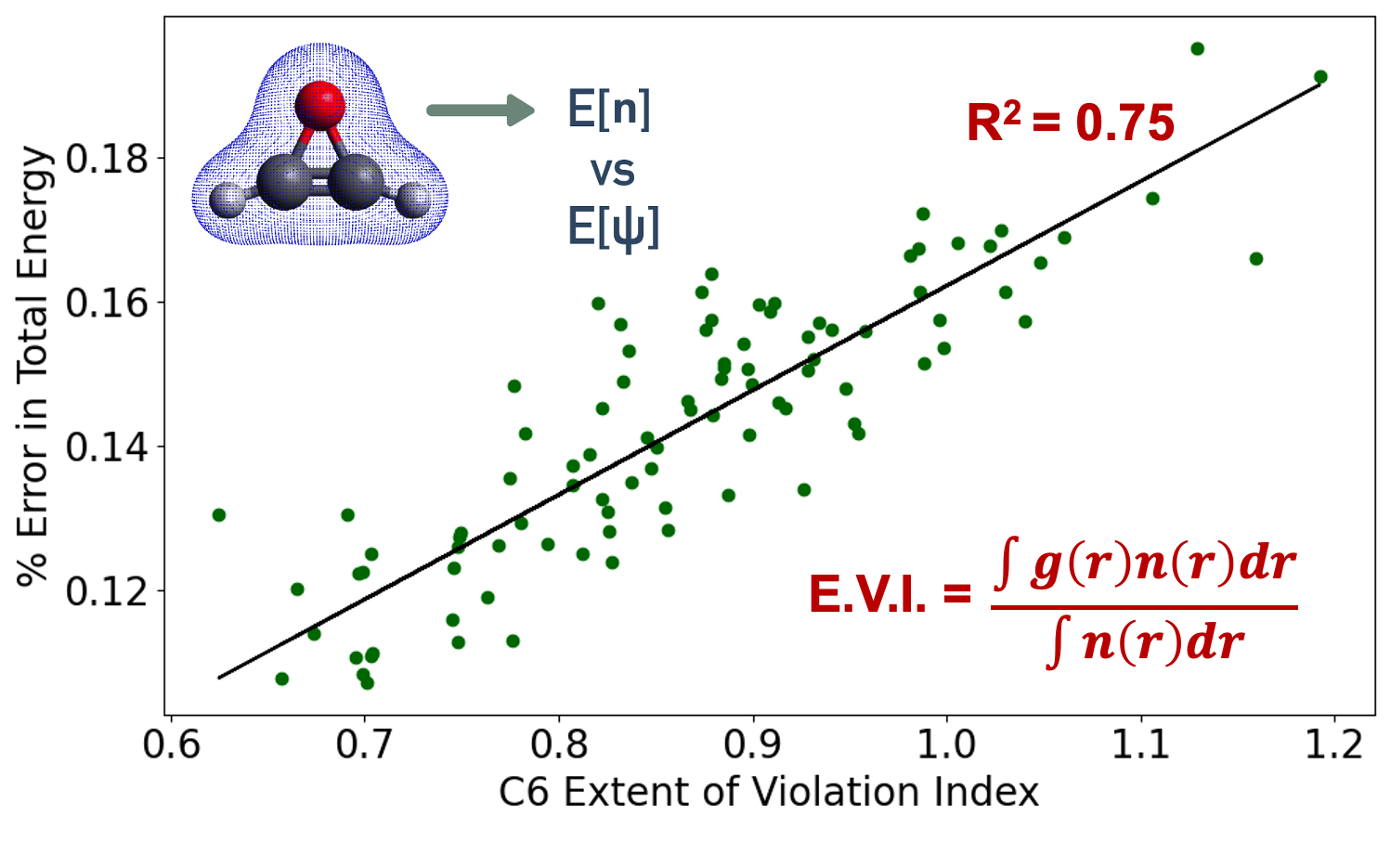}
    \end{center}
    \label{TOC Graphic}
\end{figure}

\begin{abstract}

  This work introduces Extent of Violation Indices (EVIs), a novel metric for quantifying how well exchange-correlation functionals adhere to local conditions. Applying EVIs to a diverse set of molecules for GGA functionals reveals widespread violations, particularly for semi-empirical functionals. We leverage EVIs to explore potential connections between these violations and errors in chemical properties. While no correlation is observed for atomization energies, a link emerges between EVIs and total energies. Similarly, the analysis of reaction energies suggests weak positive correlations for specific conditions, but definitive conclusions about error cancellation require advancements in both functional accuracy and our understanding of cancellation mechanisms. Overall, this study highlights EVIs as a powerful tool for analyzing functional behavior and adherence to local conditions, paving the way for future research to fully elucidate the impact of violations on energy errors.

\end{abstract}

\section{Introduction}

Density Functional Theory (DFT) has become an indispensable tool used extensively by chemists, physicists and material scientists\cite{DFT_1}. Under the Kohn-Sham ansatz\cite{Kohn_Sham} a set of non-interacting single-particle states are generated to represent the electron density. Most components of the DFT energy are known, except for the exchange-correlation term $E_{xc}$. 

\begin{equation}
    E[n(\textbf{r})] = T_s + \dfrac{1}{2}\int\int\dfrac{n(\textbf{r}_1)n(\textbf{r}_2)}{|\textbf{r}_1-\textbf{r}_2|}d\textbf{r}_1d\textbf{r}_2 - \sum_k^{nuclei}\int\dfrac{Z_k}{|\textbf{r}-\textbf{R}_k|}n(\textbf{r})d\textbf{r} + E_{xc}[n(\textbf{r})]
    \label{eqn:DFT_E}
\end{equation}

\begin{equation}\label{KS_kin}
    T_s = -\dfrac{1}{2}\sum_i^N \langle\phi_i|\nabla_i^2|\phi_i\rangle
\end{equation}
\begin{equation}
    n = \sum_{i=1}^N\langle\phi_i|\phi_i\rangle
\end{equation}
The term  $E_{xc}[n(\textbf{r})]$ in equation \ref{eqn:DFT_E}, referred to as the exchange-correlation functional, encompasses the corrections to the kinetic energy that arise due to the interacting nature of electrons and all non-classical components of the Coulomb energy. Even though the exact form of this functional remains unknown, several approximations have been developed over time that has given rise to a range of exchange-correlation functionals\cite{Gordon_Exc}.

While the exact exchange-correlation functional remains elusive, its analytical properties have guided and continue to guide functional development\cite{Constraint_review}. These analytical properties are referred to as exact conditions. Functionals that were constructed to satisfy a number of these exact conditions are generally referred to as non-empirical, for instance SCAN\cite{SCAN}, PW91\cite{PW91_1,PW91_2} and PBE\cite{PBE_1,PBE_2}. These functionals are among the best choices for solid state DFT calculations\cite{periodic_PBE}. Recently, Pederson and Burke\cite{Exact_Conditions_Pederson} studied six exact conditions by providing local expressions of these  exact  conditions, which are sufficient conditions, and evaluating them across a range of densities. They found that for some densities, semi-empirical functionals violated local conditions, and showed non-empirical functionals satisfied the conditions within a numerical threshold. 

Table \ref{tab:conditions} lists the six conditions along with their local versions. Each condition is inspired by a combination of physical and mathematical aspects, with an eye on the universality of the behavior of electron densities. That is, the conditions are independent of the external potential, and therefore are applicable to a wide range of real electron densities. Motivation for these conditions is briefly provided here, though the reader should see the original sources for further justification\cite{C2_1_C5_1,C2_2,C3_1,C4_1,C4_2,C6_1} . The first condition of Table \ref{tab:conditions} states that the DFT correlation energy should be non-positive, a concept consistent with the stabilizing effect of correlation. Since electron correlation reduces Coulomb repulsion amongst electrons, nonpositivity of $E_{c}$ is justified. While enforcing the local version of this condition ($\epsilon_{c}[{n}](\textbf{r}) \leq 0$) is not strictly necessary, it is clearly useful in the context of semi-local functionals and guarantees the  global  condition is met. Another useful concept comes from the universality of external potentials (conditions 3 and 4), where the corresponding local conditions involve the correlation enhancement factor, $F_{c}$, and its derivative with respect to the Wigner-Seitz radius, $r_{s}$\cite{Exact_Conditions_Pederson}. These are derived by stretching the electron density in space, where the exact density functional would display specific scaling inequalities in the correlation energy $E_{c}$ and the kinetic contribution to the correlation energy $T_{c}$.\cite{C3_1, C4_1, C4_2} Additional conditions coming from upper and lower bounds on the exchange-correlation potential energy ($U_{xc}$) can also be applied \cite{C2_1_C5_1}. Lower bounds have been derived from the uniform electron gas, which underpin conditions 2 and 5. These conditions are typically referred to as Lieb-Oxford bounds, and involve the Lieb-Oxford constant, $C_{LO}$, whose value is taken as 2.27\cite{Exact_Conditions_Pederson}. 
Condition 6 is closely related to the adiabatic connection \cite{Adiabatic_Connection} where the exchange-correlation energy is given as: $E_{xc} = \int_{0}^{1} U_{xc}^{\lambda}d\lambda $. In this expression, $\lambda$ is the coupling constant, which goes from 0 to 1, reflecting a transition from a system of non-interacting electrons to a system of interacting electrons. $U_{xc}^{\lambda}$ is the  exchange-correlation energy at intermediate coupling strength $\lambda$. This formula effectively links the non-interacting Kohn-Sham reference system with the fully-interacting system through a sequence of partially interacting systems, all of which share the same density. Focusing only on the correlation contribution, as $\lambda$ increases, it is imperative that the correlation energy decreases. This gives rise to the monotonicity condition for $U_{c}$, which reduces the occurrence of unphysical behaviors in the correlation energy dynamics\cite{C6_1}.

\bgroup
\def\arraystretch{1.5}%

 \begin{table}
    \centering
\caption{Exact conditions in DFT and their local counterparts.}
\label{tab:conditions}
    \begin{tabular}{|l|l|c|c|} \hline 
           \#&\multicolumn{2}{|c|}{Exact Conditions}&  Local Conditions\\\hline 
           1&$E_{c}$ non-positivity\cite{Exact_Conditions_Pederson}&$E_{c}[{n}] \leq 0$& $\epsilon_{c}[{n}](\textbf{r}) \leq 0$  \\\hline 
          2&$E_{xc}$ lower bound\cite{Exact_Conditions_Pederson,C2_1_C5_1, C2_2}&$E_{xc}[{n}] \geq C_{LO}\int d\textbf{r}\,n(\textbf{r})\epsilon _{x}^{unif}[{n}](\textbf{r})$& $F_{xc} \leq C_{LO}$  \\ \hline 
          3&$E_{c}$ scaling inequality\cite{Exact_Conditions_Pederson,C3_1}&$(\gamma - 1)E_{c}[{n_{\gamma}}] \geq \gamma (\gamma - 1)E_{c}[{n}]$& $\frac{\partial F_{c}}{\partial r_{s}} \geq 0$  \\ \hline 
          4&$T_{c}$ upper bound\cite{Exact_Conditions_Pederson, C4_1, C4_2}&$T_{c}[{n_{\gamma}}] \leq -\gamma (\frac{\partial E_{c}[{n_{\gamma}}]}{\partial \gamma }  \big|_{\gamma\rightarrow 0}) + E_{c}[{n_{\gamma}}]$& $\frac{\partial F_{c}}{\partial r_{s}} \leq \frac{F_{c}(r_{s}\rightarrow \infty )-F_{c}}{r_{s}}$  \\ \hline 
          5&$U_{xc}$ lower bound\cite{Exact_Conditions_Pederson, C2_1_C5_1}&$U_{xc}[{n}] \geq C_{LO}\int d\textbf{r}\,n(\textbf{r})\epsilon _{x}^{unif}[{n}](\textbf{r})$& $F_{xc} + r_{s}\frac{\partial F_{c}}{\partial r_{s}} \leq C_{LO}$  \\ \hline 
          6&$U_{c}(\lambda)$ monotonicity from&$\frac{\mathrm{d} U_{c}(\lambda)}{\mathrm{d} \lambda} \leq 0$& $\frac{\partial }{\partial r_{s}}\left(r_{s}^{2}\frac{\partial F_{c}}{\partial r_{s}} \right ) \geq 0$  \\ 
 & adiabatic connection\cite{Exact_Conditions_Pederson, C6_1}& &\\  \hline
    \end{tabular}

\end{table}
\egroup

An alternative category of functionals, commonly referred to as semi-empirical, holds a significant position in molecular DFT methods. As the name suggests, these functionals are inherently parameterized by fitting to benchmark datasets. Functionals such as BLYP\cite{B88,LYP_1,LYP_2}, BP86\cite{B88,P86}, M06-L\cite{M06-L_1, M06-L_2}, and B3LYP\cite{B3LYP} fall in this category. These semi-empirical functionals are designed to capture the chemical properties of molecular systems by best fits to benchmark results from accurate wavefunction computations\cite{Gordon_Exc}. However, it is important to recognize that even with a sophisticated functional form and access to extensive training data, these functionals may face challenges when applied to systems that lie outside their original training domain, such as solids\cite{Exact_Conditions_Pederson, Solid, SOLID_2}. This highlights the ongoing need to create functionals that not only excel within their training domains but also exhibit enhanced generalization capabilities to handle a wider array of chemical species and materials. In addition, many density functionals struggle to describe transition metals, which have electronic structures that differ considerably from the main-group elements that dominate DFT training sets\cite{DFT_TM,DFT_TM2,DFT_TM3,DFT_TM4,DFT_TM5,Mn2Si12, Rask_TMC1, Rask_TMC2}. 

An exact exchange-correlation functional is anticipated to satisfy all exact conditions, including but not limited to those present in Table \ref{tab:conditions}. As a consequence, one might expect non-empirical functionals to outperform their empirical counterparts, especially when applied across wide chemical and materials spaces. 
In the realm of molecular calculations, however, semi-empirical functionals have maintained substantial practical value.\cite{Datasets, DFT_R1, DFT_R2, Gordon_Exc, dft_electro} In light of the apparent success of semi-empirical functionals in predicting chemical properties, one might ask: is there any correlation between satisfying exact conditions and predicting chemical properties?

The above question might be addressed, at least to a certain degree, through metrics that measure the degree of violation of exact conditions. An exact functional would adhere precisely to all known exact conditions, and most likely the degree of violation of each exact condition would correlate to errors in property evaluation. No present-day functional is anywhere near exact, so it is unclear whether the same assumption applies to contemporary functionals. This leads us to analyze contemporary approximate density functionals and seek relationships between exact conditions and energetic properties. The methods section introduces an index that measures the extent of violation, averaged over a molecule's electron density, which may be useful in finding correlations between local exact conditions and electronic energies. 

\begin{figure}[h]
    \centering
    \includegraphics[width=\textwidth]{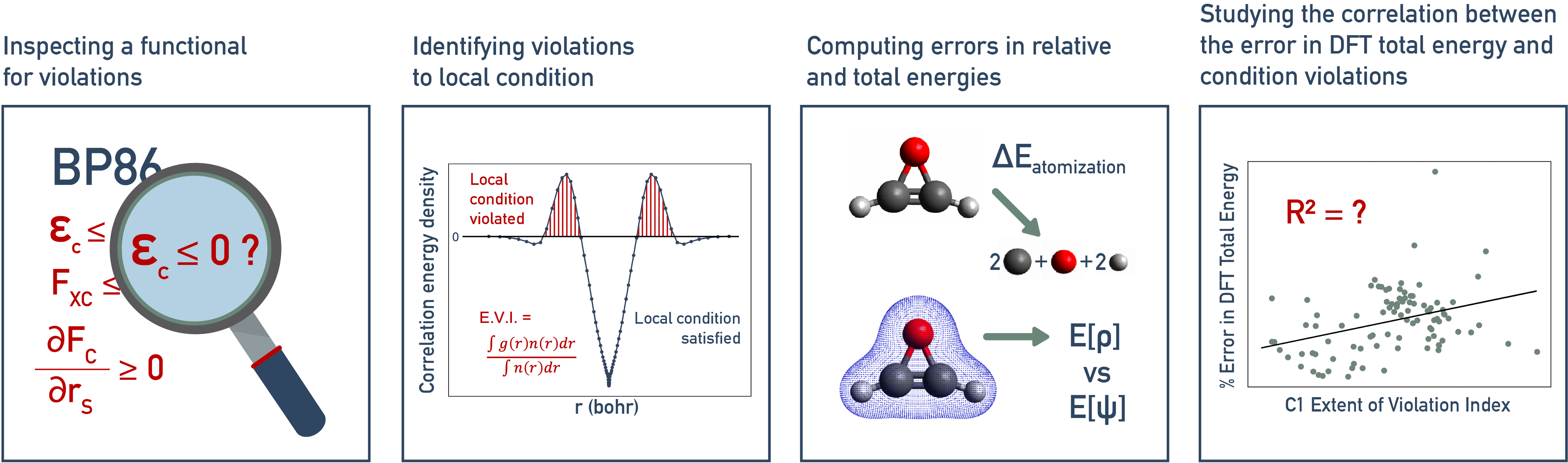}
    \caption{Exact condition analysis workflow: Close inspection of a functional reveals violation of a local condition. Errors in atomization and DFT total energies are calculated, and the relationship between the error in DFT total energy and condition violations is explored.}
    \label{Workflow}
\end{figure}

This study examines a selection of non-empirical and semi-empirical generalized gradient approximation (GGA) exchange-correlation functionals. Our investigation assesses these functionals for potential deviations from local conditions for a diverse range of molecules (Figure \ref{Workflow}). Errors in total and relative energies predicted by these functionals are related to the extent to which a functional adheres to these conditions, demonstrating that the new violation index is a useful means for examining approximate density functionals.

\section{Methods}

In order to investigate the chemical significance of local conditions, we studied the impact of violating these conditions on predictions of two chemical properties, namely atomization and reaction energies. Reliable values of these properties were obtained from two databases, namely W4-11 and G2RC, which are often used in DFT benchmark studies\cite{W4-11, G2RC_1, G2RC_2}. We restricted our study to GGA density functionals, since they form the simplest semi-local models and are used extensively. In order to compare the extent to which different functionals violate a given condition, we quantified violations by computing the extent of violation index, which is defined later in this section.

In this study, we selected closed-shell neutral molecules from the W4-11 and G2RC databases. This selection was prompted by the fact that closed-shell neutral molecules exhibit simplified electronic structures, avoiding strong correlations where DFT is generally less precise.

\begin{figure}[h]
    \centering
    \includegraphics[width=\textwidth]{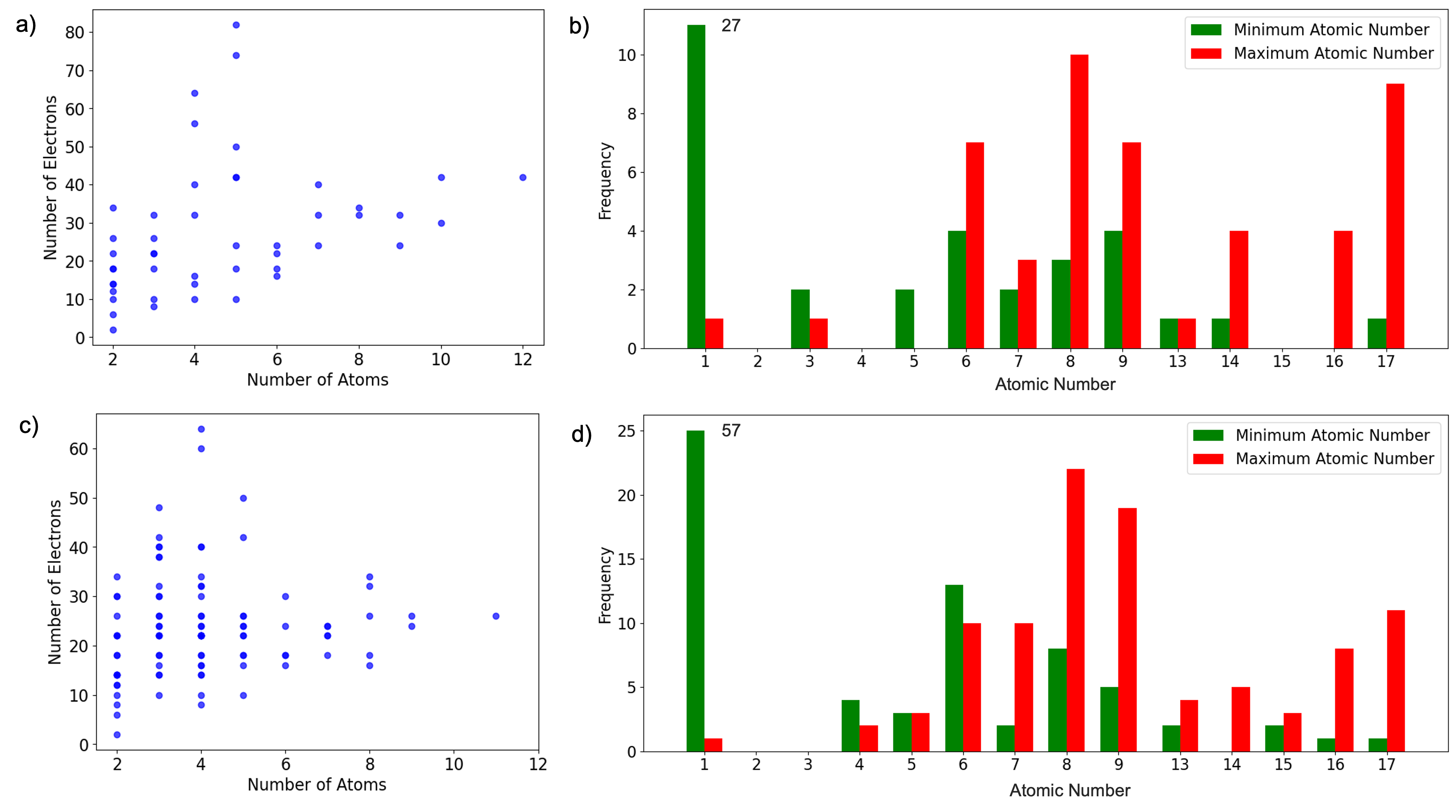}
    \caption{Plots summarizing the properties of molecules in the G2RC (a and b) and W4-11 (c and d) databases. The frequency values have been clipped to 11 and 25 in plots b and d.}
    \label{Datasets}
\end{figure}

Figure \ref{Datasets} delineates chemical space that is represented in the two datasets.
The W4-11 database contains a diverse range of 140 molecules, from diatomics to medium-sized organic compounds. These molecules predominantly consist of atoms from the second row of the periodic table, although some also incorporate atoms from the third row, and a few have a combination of both. On the other hand, the G2RC database, while similar in elemental composition, features a smaller number of molecules, providing reaction energies for 25 reactions.

To generate electron densities for the selected molecules, we employed very dense (PySCF level 9) Becke atomic grids (200 radial and 1454 angular grid points) \cite{grid1} using the PySCF program \cite{PySCF1, PySCF2}. All computations used the augmented, polarized, triple-zeta basis set, aug-cc-pVTZ\cite{augccpvtz}. 
We also calculated errors in DFT total energies of molecules using PySCF, with the CCSD(T) method serving as the ground truth. To evaluate DFT energies and densities, expressions for semi-empirical GGA functionals (BLYP, BP86, OLYP\cite{OPTX_OLYP_exchange, LYP_1, LYP_2}, SOGGA11\cite{SOGGA11}, GAM\cite{GAM}, N12\cite{N12}), non-empirical GGAs (PBE, PW91, AM05\cite{AM05_1, AM05_2}) and their derivatives with respect to the density were obtained from the LibXC library\cite{LibXC}. All source code for the calculations described in this work can be found on our group's GitHub page at \url{https://github.com/ZimmermanGroup/Local_Conditions_DFT}.

In order to evaluate local conditions, the correlation energy density $\epsilon_{c}[{n}](\textbf{r})$  for each functional was obtained from the LibXC library.\cite{LibXC} The exchange-correlation enhancement factor was computed as $F_{xc}$ = $\epsilon_{xc}[{n}](\textbf{r})$/$\epsilon_{x}^{unif}[{n}](\textbf{r})$ where $\epsilon_{x}^{unif}$ is the exchange energy density for an unpolarized uniform electron gas and is given as $\epsilon_{x}^{unif}$ = $-(3/4\pi)(3\pi^2n)^{1/3}$. 
The Wigner-Seitz radius was computed as $r_s$ = $(4\pi n/3)^{-1/3}$. The derivative $\frac{\partial F_{c}}{\partial r_{s}}$ was calculated by substituting $F_c$ as  $\epsilon_{c}[{n}](\textbf{r})$/$\epsilon_{x}^{unif}[{n}](\textbf{r})$ and making use of the quotient rule: \begin{equation}
    \frac{\partial F_{c}}{\partial r_{s}} = \frac{\epsilon_{x}^{unif}\frac{\partial \epsilon_{c}}{\partial r_{s}} - \epsilon_{c}\frac{\partial \epsilon_{x}^{unif}}{\partial r_{s}}}{(\epsilon_{x}^{unif})^2}
\end{equation}

\textbf{Extent of Violation Index:}
Each local condition was evaluated across the entire grid for each molecule, establishing a distribution of values, $g(\textbf{r})$, which represents the deviation from that condition's equalities. This information is condensed into a metric to quantify the overall degree in which the exact condition is violated. This work therefore defines the Extent of Violation Index (EVI) as

\begin{equation}
    EVI = \frac{\int g(\textbf{r}) n(\textbf{r}) d\textbf{r} }{\int n(\textbf{r}) d\textbf{r}}    
\end{equation}

\begin{equation}
  g(\textbf{r}) =
    \begin{cases}
      |violation| & \text{if local condition is violated}\\
      0 & \text{otherwise}
    \end{cases}       
\end{equation}
For example, while evaluating the second local condition, if at a grid point, $F_{xc} > C_{LO}$, $g(\textbf{r})$ will be equal to $F_{xc} - C_{LO}$. If, however, $F_{xc} \leq C_{LO}$, $g(\textbf{r})$ will be set to zero. Numerical integral over the grid gives the extent of violation. While violating local conditions does not strictly imply the   global  exact conditions are violated, evaluation of the local conditions is useful for assessing semi-local functionals nonetheless\cite{Exact_Conditions_Pederson}.

The EVI was calculated for every molecule in the two databases and all nine GGA functionals considered in this study.
We also computed EVI for reactions in the G2RC dataset. The violation index for a reaction was computed as follows. Consider the following reaction: \ce{A + B -> C + D}. For any local condition $C_m$, we computed the EVI for this reaction as:

\begin{equation}
    EVI_{reaction}^{C_m}\; = \; N_{C} \:EVI_{C}^{C_m} \;+\; N_{\ce{D}} \:EVI_{\ce{D}}^{C_m} \;-\; N_{\ce{A}} \:EVI_{\ce{A}}^{C_m} \;-\; N_{\ce{B}} \:EVI_{\ce{B}}^{C_m}
\end{equation}
Multiplication of EVIs of molecules with their number of electrons ($N_i$) was required since EVIs for each individual molecule were normalized. 

Our violation indices differ from the metric computed by Pederson and Burke for evaluating local conditions\cite{Exact_Conditions_Pederson}. In their work, Pederson and Burke constructed a range of electron densities and their gradients over uniform spacing between realistic limits. Relevant derivatives involved in local conditions were then computed numerically. They reported the fraction of grid points where local conditions were violated, where violations were computed using predefined tolerance values. 
In comparison, our metric includes the magnitude of violation as well as weighting (and averaging) by the density. While similar in spirit to the metric of Ref. \cite{Exact_Conditions_Pederson} additional concepts will be revealed by the EVI metric used herein.

\section{Results}
Nine GGA exchange-correlation functionals were examined for their adherence to the local conditions of Table \ref{tab:conditions}. First to be examined are non-empirical functionals, which are built to satisfy a number of exact conditions. For example, the non-empirical functional PBE by construction satisfies several energetically significant conditions, such as correlation energy non-positivity (Condition 1 in Table \ref{tab:conditions}), Lieb-Oxford bounds (Conditions 2 and 5), uniform scaling to the high-density limit for the correlation energy, uniform density scaling for exchange energy, the exact exchange energy spin-scaling relationship, and the linear response of the spin-unpolarized uniform electron gas\cite{PBE_1,PBE_2}. Therefore Conditions 1, 2, and 5 should be strictly adhered to in PBE, though Conditions 3, 4, and 6 could be violated.  Next, the local conditions for semi-empirical functionals will be examined. These functionals were primarily designed to predict chemical properties by fitting to benchmark results involving molecular systems. An example of this category is the SOGGA11 functional, that has a flexible functional form which satisfies two physical constraints (the uniform electron gas limit and the second order density-gradient expansion) and has 18 free parameters that are optimized by fitting to 15 chemical databases. \cite{SOGGA11} For each category of functionals, the relationship between exact local conditions and their impact on successful predictions of chemical properties will be studied.

\subsection{Non-empirical functionals}

\begin{figure}[h]
    \centering
    \includegraphics[width=0.5\textwidth]{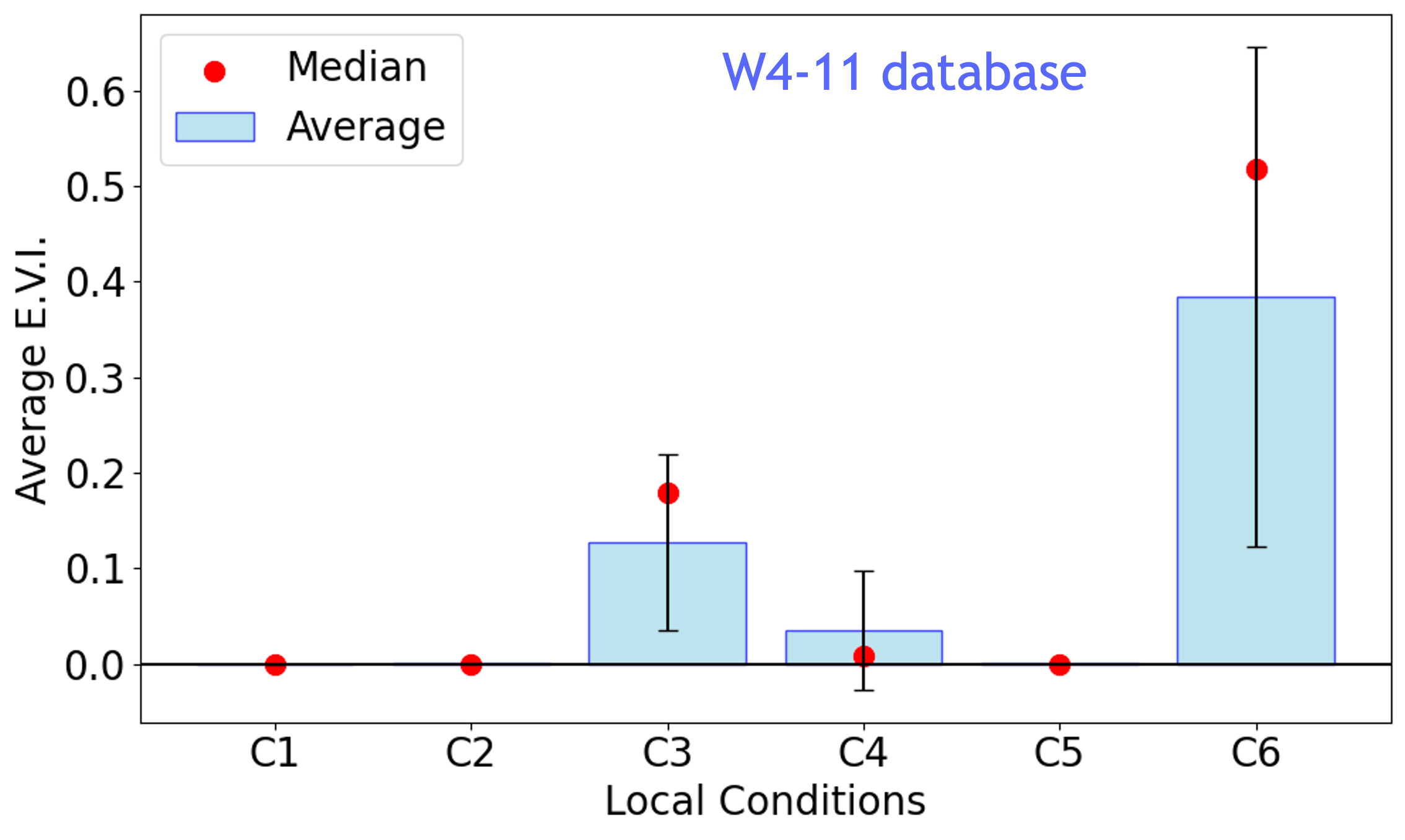}
    \caption{Distribution of extent of violation indices for all local conditions for non-empirical functionals PBE, PW91 and AM05 (W4-11 database). For each local condition, the median value, average and standard deviation of EVIs is displayed.}
    \label{Non_emp}
\end{figure}

First, we investigate non-empirical GGA functionals for violations to local conditions. For the three non-empirical functionals, namely PBE, PW91 and AM05, Figure \ref{Non_emp} shows the distribution of EVIs for all local conditions. EVIs are computed over the set of neutral closed shell molecules in the W4-11 database for each functional, then the distributions over these values are in Figure \ref{Non_emp}. Since these functionals were constructed\cite{PBE_1,PBE_2, PW91_1,PW91_2, AM05_1, AM05_2} to satisfy conditions 1, 2, and 5, there are no violations detected in those categories ($E_{c}$ non-positivity, $E_{xc}$ lower bound, and  $U_{xc}$ lower bound local conditions). The average extent of violation indices calculated for conditions 3, 4, and 6 — conditions not explicitly considered during the development of these functionals — show non-zero values. 

\subsection{Exact Conditions and Chemical Properties}
Investigation of semi-empirical functionals (BLYP, OLYP, BP86, SOGGA11, N12, and GAM) revealed violations to all local conditions, albeit to varying extents. The average violation indices for 96 closed shell molecules in the W4-11 database are reported in Table \ref{tab:scores} for all local conditions.

In order to better understand magnitudes of extent of violation indices, we look at the range of values of quantities that appear in local conditions. These are reported for He using the BLYP functional. \emph{Condition 1}: The correlation energy density values range from -0.05 a.u. close to the nucleus, to 0.02 away from it. The average EVI values for the $E_c$ non-positivity condition in Table \ref{tab:scores} are two orders of magnitude smaller than the most positive correlation energy density value (0.0002 vs 0.0217). \emph{Conditions 2 and 5}: The exchange-correlation enhancement factor, which is the dominant term in local conditions 2 and 5, has values close to 1 in the vicinity of the He atom. Large values of the enhancement factor, that exceed the Lieb-Oxford constant (taken to be 2.27), are seen at large distances from the nucleus, where there is little electron density. Much larger values (around 490 a.u.) are also encountered at very large distances due to diminishing values of the denominator of the enhancement factor ($\epsilon_{x}^{unif}$ = $-(3/4\pi)(3\pi^2n)^{1/3}$, \ce{n -> 0}). \emph{Condition 3}: The values of $\frac{\partial F_{c}}{\partial r_{s}}$ range from -0.57 to 0.08. Since EVI values for all semi-empirical functionals studied here  exceed 0.08 (which is the maximum value of $\frac{\partial F_{c}}{\partial r_{s}}$ for He)  for condition 3, these functionals exhibit significant violations to the $E_c$ scaling inequality (C3). \emph{Condition 6}: Along similar lines, we conclude that these functionals also show large violations to the $U_c$ monotonicity condition (C6).

\begin{figure}[h]
    \centering
    \includegraphics[width=\textwidth]{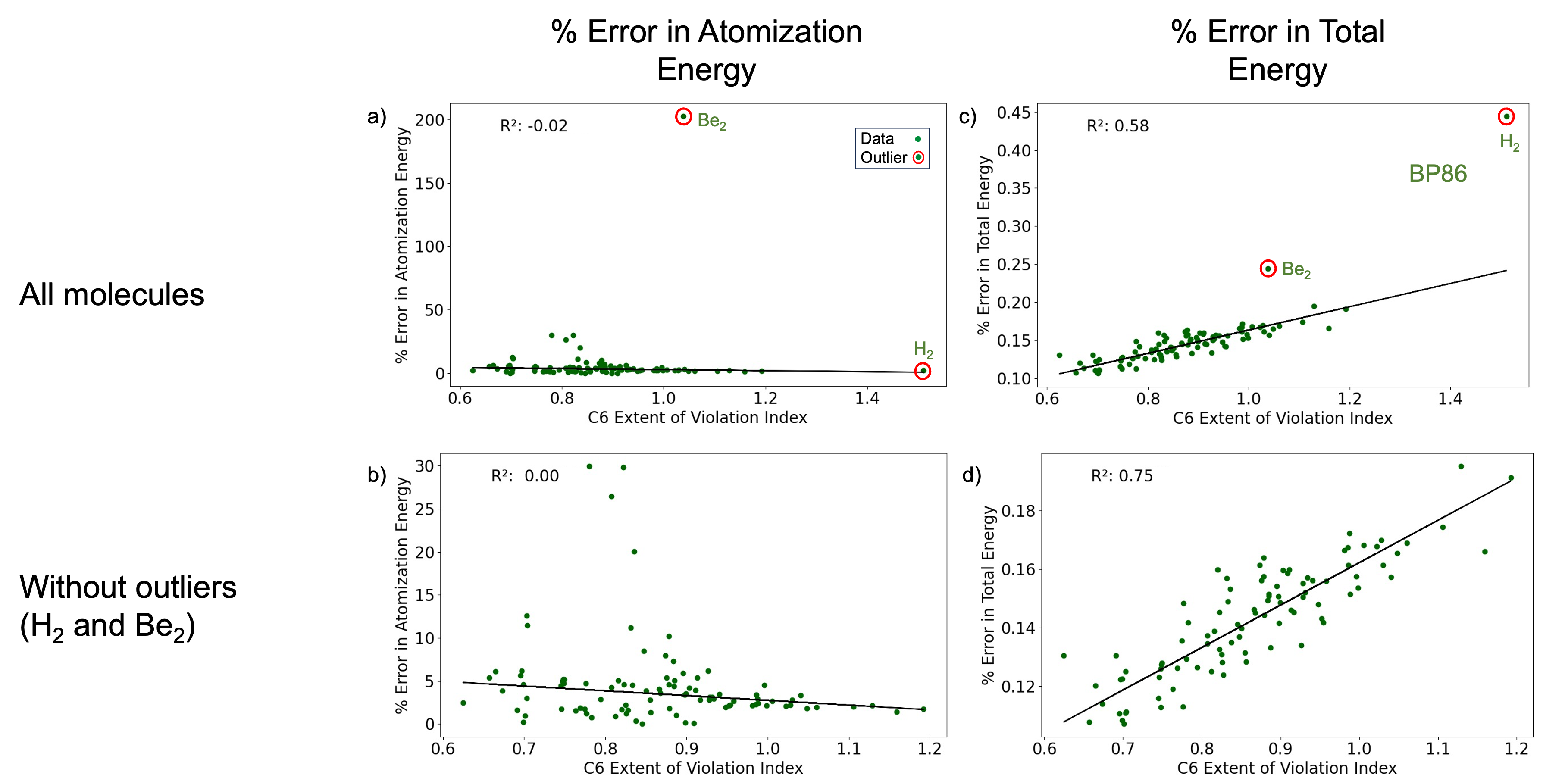}
    \caption{Variation of \% error in atomization energy (a, b) and \% error in total energy (c, d) with the extent of violation index for local condition 6 for BP86 functional. Figures b and d exclude molecules \ce{H2} and \ce{Be2} (W4-11 database).}
    \label{Atom_vs_total}
\end{figure}

The EVI scores of Table \ref{tab:scores} indicate statistically notable deviations from exact local conditions, but do not indicate precisely how they might affect chemical properties. To understand the relationship between the extent of violation index values and chemical property predictions, we studied the correlation between these scores and the errors in atomization energies for all 96 closed-shell neutral species in the W4-11 database. As an example of typical results, Figure \ref{Atom_vs_total} a) shows the variation of percent error in atomization energy with the extent of violation index for the $U_{c}$ monotonicity condition (C6) for BP86. There appears to be no correlation between the two quantities, as evident from an $R^2$ value close to zero. Repeating this exercise with all other conditions yields the same result: the error in atomization energy is found to be insensitive to EVI. 

\begin{table}
    \centering
    \begin{tabular}{|c|c|c|c|c|c|c|} \hline 
         Functional&  C1 average&  C2 average&  C3 average&  C4 average&  C5 average& C6 average\\
 & EVI& EVI& EVI& EVI& EVI&EVI\\ \hline 
         BLYP&  0.0002&  0.0025&  0.0855&  0.0295&  0.0020& 1.4852\\ \hline 
         OLYP&  0.0002&  0.0006&  0.0855&  0.0295&  0.0003& 1.4852\\ \hline 
         BP86&  0.0000&  0.0029&  0.1729&  0.0010&  0.0070& 0.8739\\ \hline 
         SOGGA11&  0.0000&  0.0192&  0.0805&  0.3768&  0.2288& 2.5847\\ \hline 
         N12&  0.0000&  0.0030&  0.1010&  0.0705&  0.1279& 1.4819\\ \hline 
         GAM&  0.0008&  0.0000&  0.0924&  9.1014&  0.0003& 2.7285\\ \hline\hline
 Range& 0.0000 -& 0.0000 -& 0.0805 -& 0.0010 -& 0.0003 -&0.8739 -\\
 & 0.0008& 0.0192& 0.1729& 9.1014& 0.2288&2.7285\\\hline\hline
    \end{tabular}
    \caption{Average extent of violation indices for all local conditions for semi-empirical functionals, reported for closed shell neutral molecules in the W4-11 database.}
    \label{tab:scores}
\end{table}

\subsection{Exact Conditions and DFT Total Energies}
Having examined the relationship between EVI and atomization energies, the total energy was considered next. For the closed-shell molecules studied herein, CCSD(T) provides excellent total energies as benchmark values.\cite{CCSDT_1, CCSDT_2}
Hence, we computed the difference between total energies predicted by semi-empirical functionals and CCSD(T) total energies. We consider this difference as the error in DFT total energy, and study its correlation with the EVI for conditions with significant violations. Figure \ref{Atom_vs_total} c) shows the variation of percent error in total energy with the extent of violation index for the $U_{c}$ monotonicity condition (C6) and the BP86 functional for molecules in the W4-11 database. Significant correlation between the two quantities was indicated by the $R^{2}$ value of 0.58. 

\begin{figure}[h]
    \centering
    \includegraphics[width=0.7\textwidth]{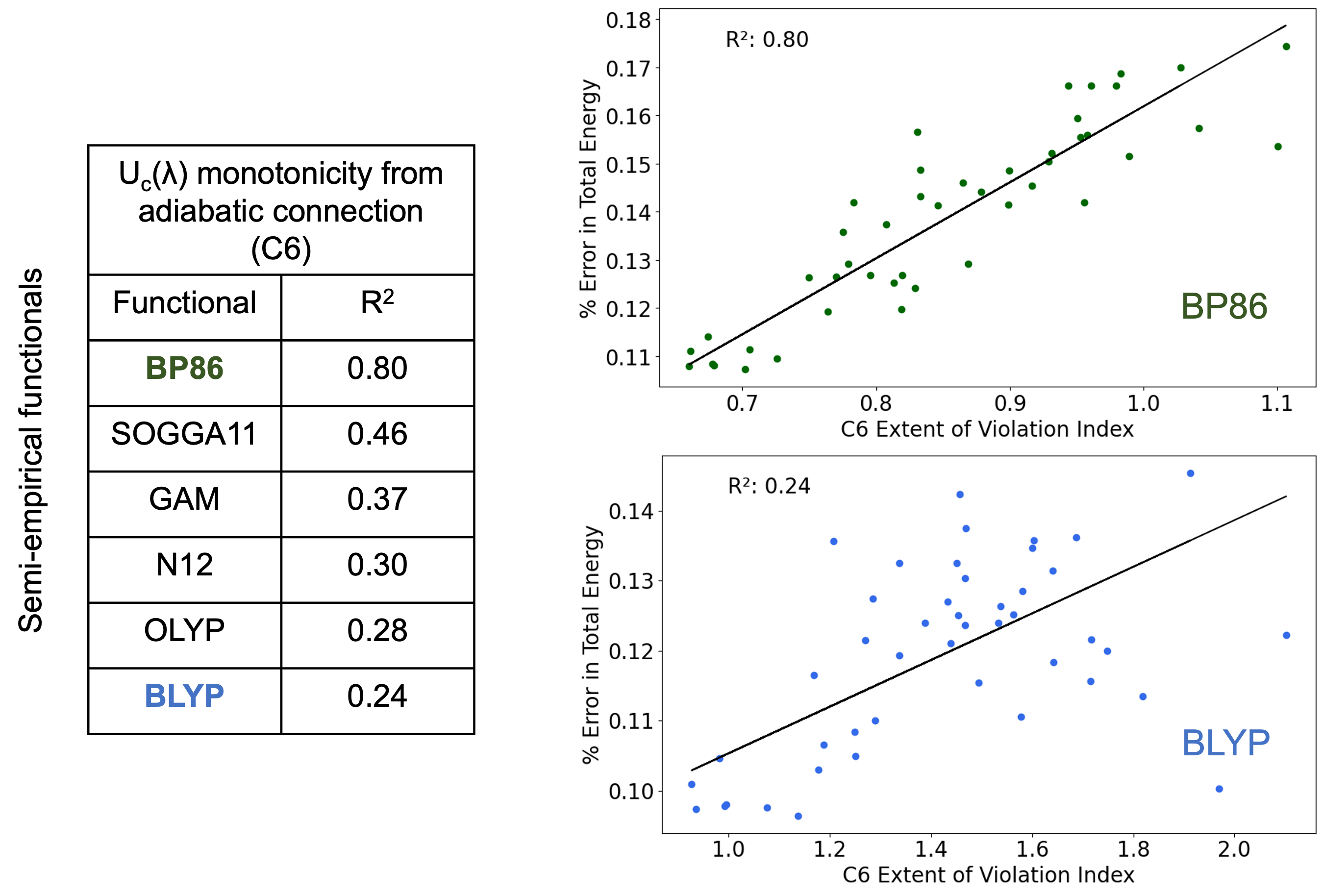}
    \caption{R\textsuperscript{2} values from the variation of percent error in total energy with the extent of violation index plots for local condition 6 for semi-empirical functionals reported for the G2RC database (excluding outliers). Shown are the BP86 and BLYP functionals, which have the highest and lowest R\textsuperscript{2} values.} 
    \label{R2_min_max}
\end{figure}

The other semi-empirical functionals considered in this study also showed the same trend, the errors in their total energy predictions correlated with their violation indices for condition 6. However, the extent of this correlation varied across different functionals. Figure \ref{R2_min_max} shows the percent error in total energy vs C6 violation index for BP86 and BLYP functionals. These plots focus on the closed-shell neutral molecules that appear in the 25 reactions in the G2RC database. BP86 shows an $R^2$ value of 0.80, and BLYP displaying a weaker trend with $R^2$ of 0.24. The figure also presents the $R^2$ values for other functionals, which exhibit substantial variation. This variation is expected, considering that these functionals, although all GGAs, were constructed in different ways, involving varying levels of data fitting: ranging from 4 parameters fit to He in LYP to 18 parameters fit to 15 chemical datasets in SOGGA11.

\begin{figure}[h]
    \centering
    \includegraphics[width=\textwidth]{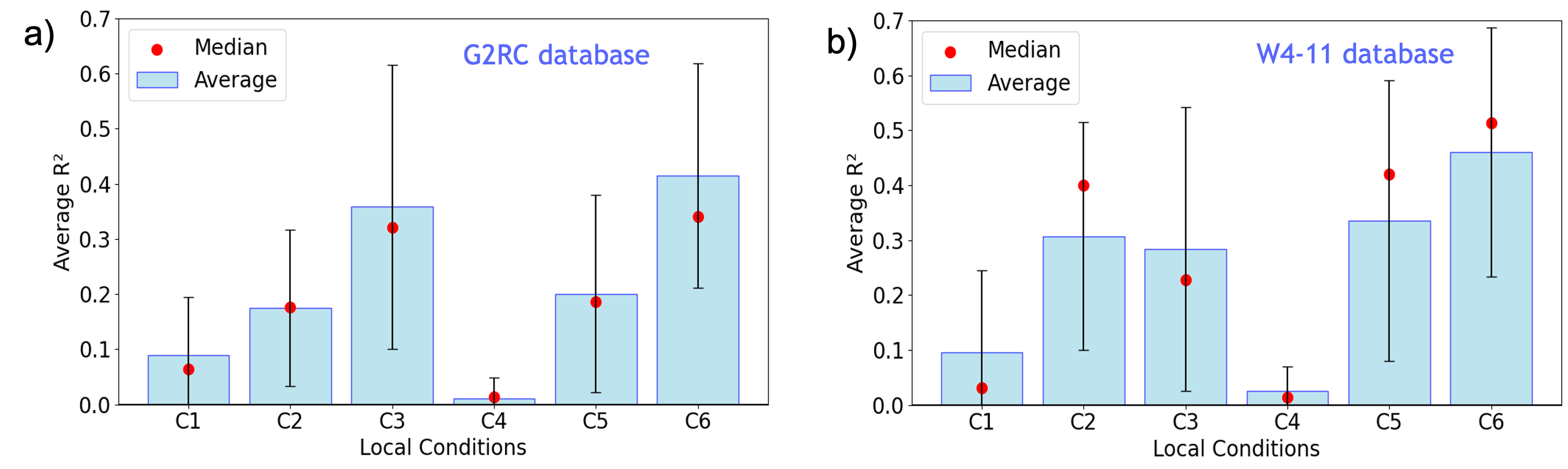}
    \caption{Average values of R\textsuperscript{2} relating EVI to total energy for semi-empirical functionals across a) G2RC and b) W4-11 databases.}
    \label{Avg_R2}
\end{figure}

In both the atomization energy and total energy plots, we saw that molecules $H_{2}$ and $Be_{2}$ were outliers. These outliers persisted in similar analyses of conditions other than Condition 6. Consequently, we recalculated the $R^2$ values without considering these outliers, as shown in Figure \ref{Atom_vs_total} b) and d). While no correlation with atomization energy was found without the outliers, the correlation between violation index and error in total energy strengthened.  $H_{2}$ was the only two-electron system considered in our study, which could explain its distinct behaviour, and $Be_2$ is unique as a near-zero bond order system. Hence, commonly used density functional approximations fail to give accurate predictions for either of these two chemical species\cite{Be2}.

So far, only violation indices for the $U_{c}$ monotonicity condition have been examined. Statistical correlations between errors in total energies and EVI are also observed for other local conditions. Figure \ref{Avg_R2} shows the average $R^2$ values for all local conditions. Notably, not every condition significantly correlates with the total energy. The $R^2$ value for the $T_{c}$ upper bound condition (C4) is nearly zero. The $E_{c}$ scaling inequality (C3) and the $U_{c}$ monotonicity condition (C6) have the largest $R^2$ overall across the two benchmark sets. The $R^2$ values for $E_{xc}$ and $U_{xc}$ lower bounds (C2 and C5), however, are nearly as large as those of C3 and C6 for the W4-11 database.

\section{Discussion}

\subsection{How Local Conditions Relate to Energies}
The analysis of EVI in Table 2 and Figures 3-5 confirmed that local conditions can be violated in semiempirical functionals, and to a lesser extent, even in nonempirical functionals. In this section we discuss the EVI metric and ask what does it tell us about the utility of the local conditions of Table 1. Before doing so, we briefly discuss the relationship of the present study to its motivating precedent. 

The recent communication by Pederson and Burke\cite{Exact_Conditions_Pederson} introduced the C1-6 local conditions and looked for violations of the same. Their analysis employed idealized Gedanken densities, contrasting with the molecular densities used in this study. Ref \cite{Exact_Conditions_Pederson} quantified violations by counting the number of violations above certain preset thresholds. In contrast, our EVIs account for the magnitude of the violations, not just their presence. This approach revealed correlations between local condition violations and errors in total energies. While both metrics offer valuable insights, they serve complementary purposes.

\begin{figure}
    \centering
    \includegraphics[width=1\linewidth]{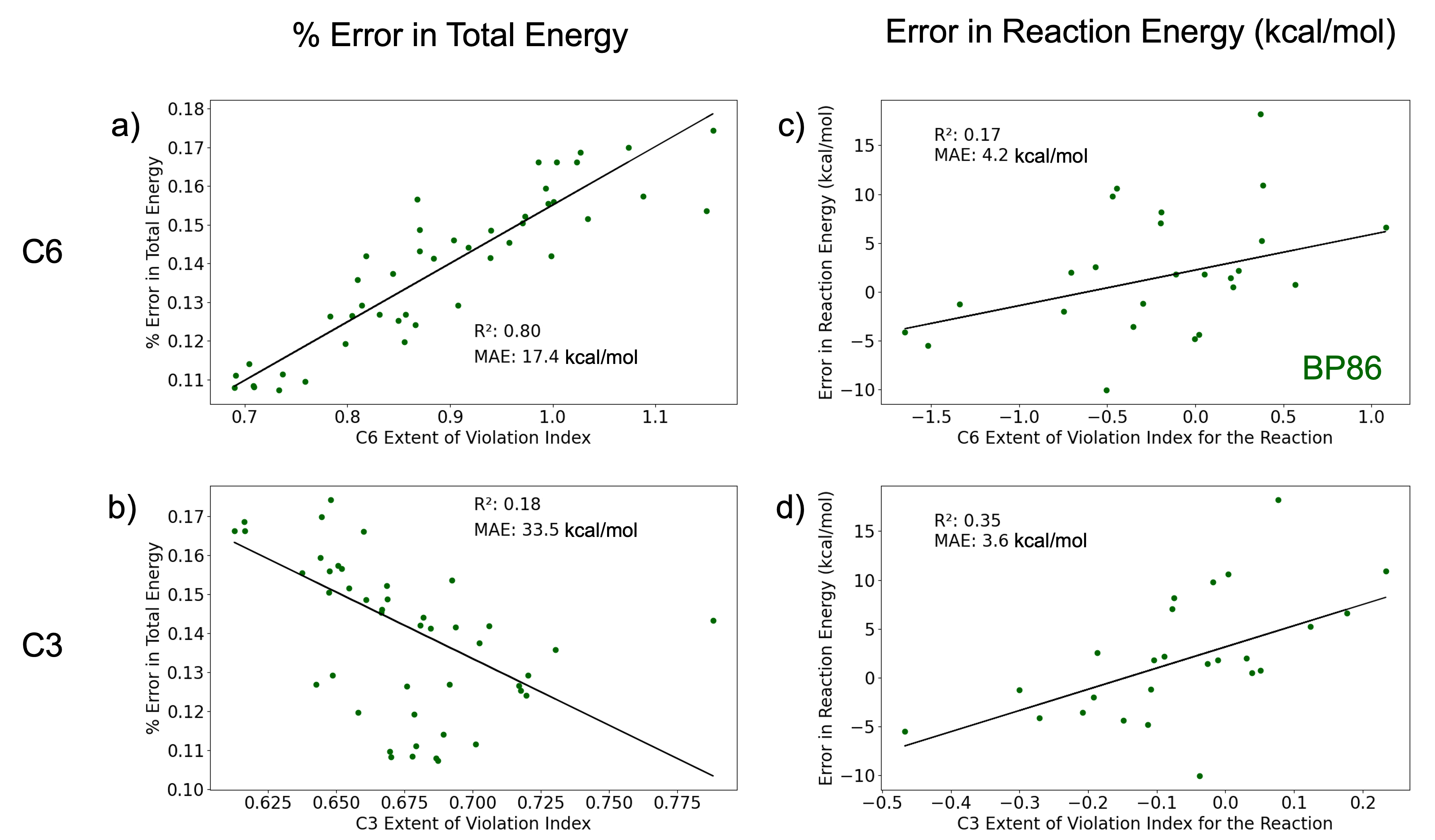}
    \caption{Variation of percent error in total energy with the extent of violation index for a) local condition 6 and b) local condition 3 for BP86 functional for molecules in the G2RC dataset. Plots c) and d) depict the variation of error in reaction energy with the  extent of violation index for the reaction for reactions in the G2RC dataset. Mean absolute errors (MAEs) in plots a) and b) were converted from percent values to kcal/mol.}
    \label{fig:EVI_react}
\end{figure}

\begin{figure}[h]
    \centering
    \includegraphics[width=0.5\textwidth]{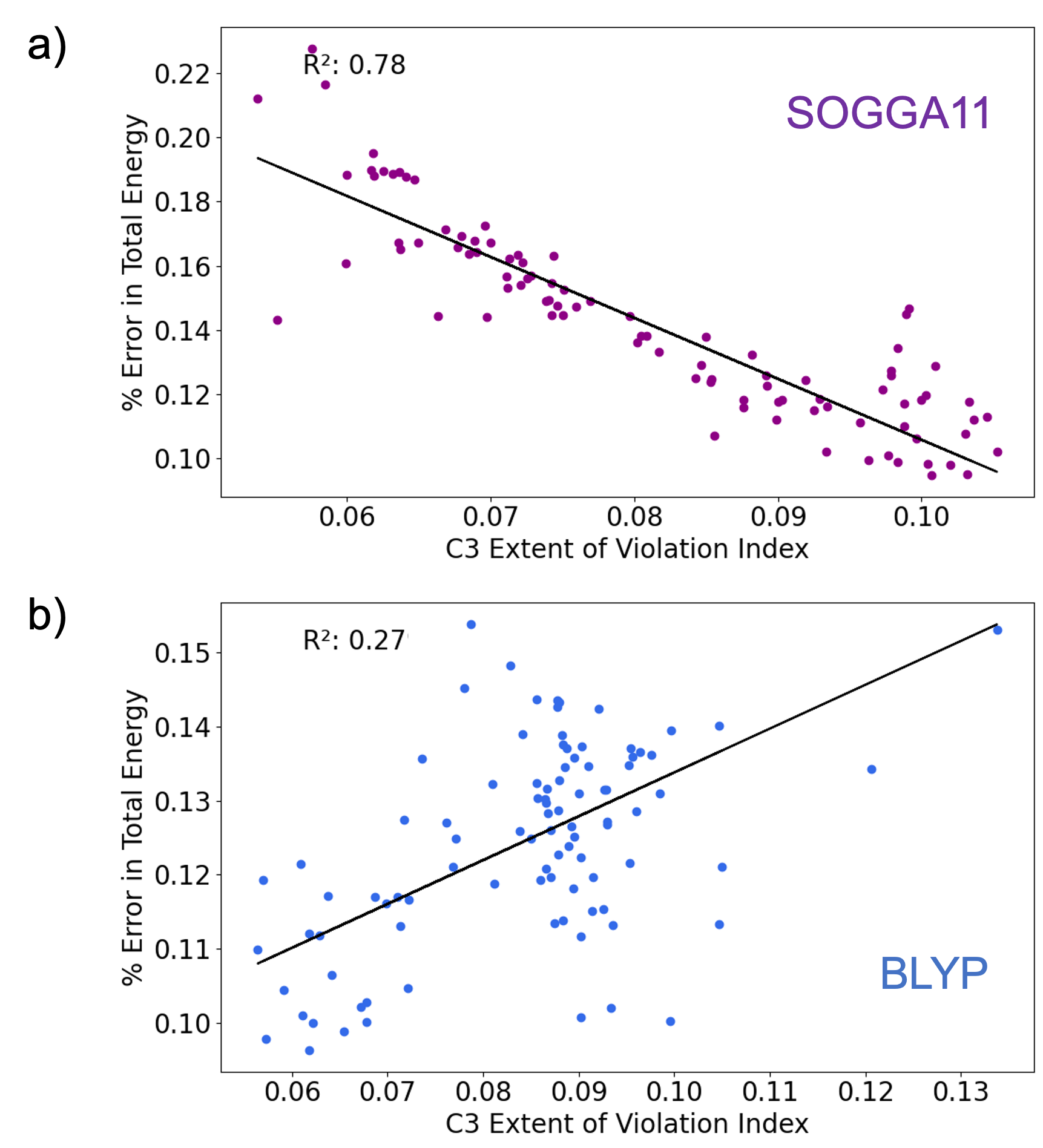}
    \caption{Variation of percent error in total energy with the extent of violation index for local condition 3 for a) SOGGA11 and b) BLYP functionals reported for the W4-11 database.}
    \label{Negative_corr}
\end{figure}

Returning to the analysis of local conditions via the present study, the lack of correlation between EVI and atomization energies contrasts with their noticeable relation to total energy for GGA functionals (the latter up to $R^2=0.80$, depending on functional and condition). For BP86, where C6 correlated at $R^2=0.80$ with the error in total energy, it is natural to ask how this factor carries over into relative energies. Since the atomization energies of Figure \ref{Atom_vs_total} are uncorrelated to C6, and also uncorrelated to C1-5, the total energy relationship with EVI appears to have no obvious effect on relative energies. A possible explanation for this is found in Figure \ref{fig:EVI_react}, which compares the C6 and C3 EVIs to the energy errors in the G2RC dataset. There, the mean absolute errors (MAEs) in total energy are much higher than the MAEs in reaction energy. Even when a substantial variation in total energy is described by a correlation with EVI, there is still a large variation that does not cancel out in a relative energy calculation. While cancellation of errors is undoubtedly present in GGA relative energies,\cite{DFT_Error1, DFT_Error2, DFT_Error4} it is not easy to pinpoint the source of error cancellation when considering the EVI metrics. 

The correlation between total energy and EVI can even be negative (Figure \ref{fig:EVI_react}b), suggesting that increased violation of local conditions can \textit{improve} total energies. While this correlation is weak for BP86 with C3 ($R^2=0.18$), a significant $R^2$ was found for a related functional. Figure \ref{Negative_corr} shows relationships between error in total energy and the EVI for C3, particularly for the SOGGA11 and BLYP functionals. These two functionals were chosen because the errors in their total energy predictions showed the strongest negative and positive correlations with the EVI for condition 3 in the W4-11 dataset. For the SOGGA11 function, increased EVI leads to improved total energies, with $R^2$ = 0.78. For this functional, amongst others, it cannot be concluded that reduction in EVI will lead to improved total energies.

\subsection{A Simple Model for Constraints}

\begin{figure}
    \centering
    \includegraphics[width=1\linewidth]{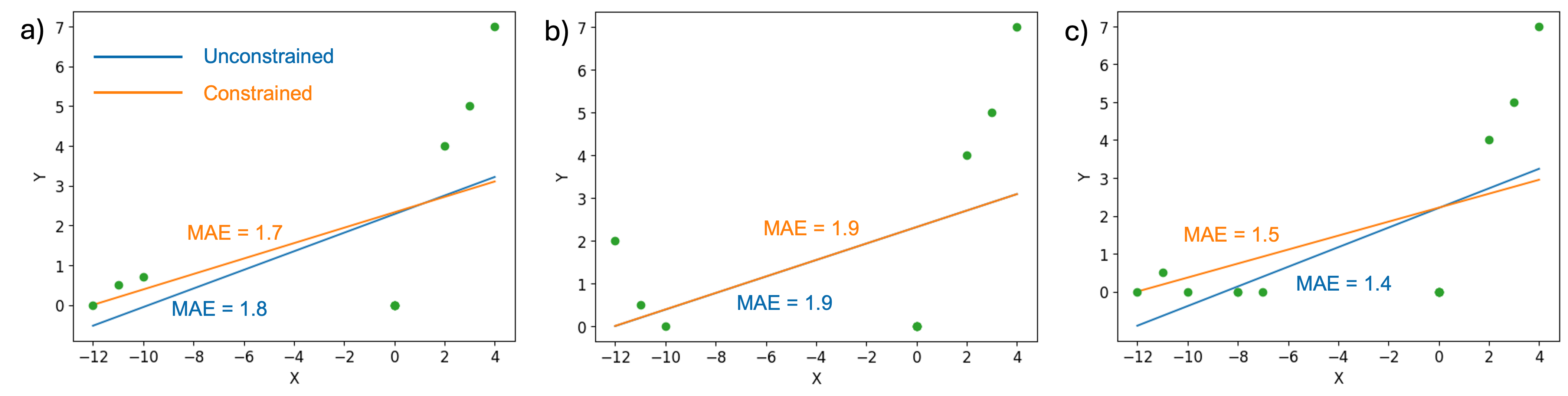}
    \caption{Regression experiment  where a linear model was fit to predict a non-negative target  based on one feature. Enforcing the non-negativity constraint led to a diverse range of outcomes: a) improved  predictions, b) no significant change and c) slightly worse predictions, as measured by the mean absolute error (MAE).}
    \label{fig:regression}
\end{figure}

We found the statistical relationships of Figure \ref{fig:EVI_react}a and Figure \ref{Negative_corr}a to be counterintuitive. While one might protest that various complications of density functional modeling could lead to this result, the same kind of behavior is possible even in very simple settings. Figure \ref{fig:regression} compares linear regression models ($y=mx+b$) for three datasets where $y \geq 0$. These datasets are entirely invented, but represent a prototypical case where a dataset is to be fit to a simple, low-dimensional model. The linear fit can be done with or without enforcing the constraint, $y \geq 0$. As can been seen, all three possibilities emerge: (a) the constrained model has lower MAE, (b) the model is the same with or without constraints, and (c) the model has higher MAE upon application of the constraint. Depending on the quality of the training data, the quality of the model, and the form of the constraint, the constrained models may or may not perform better than the baseline. 

\section{Conclusions}
This study introduced the EVI metric to gauge how far a density functional strays from exact conditions in molecular systems. The metric was applied to a number of GGA exchange-correlation functionals, showing significant statistical relationships between EVI and total energies for semi-empirical functionals. Surprisingly, the relationship could even be a negative trend, suggesting a counterintuitive possibility: increasing violations for certain local conditions (e.g., C3) might be associated with improved total energies.

Precise interpretation of these negative correlations necessitates further investigation. While various complexities inherent to density functional modeling could contribute to this result, analysis in a much simpler setting shows the same phenomenon. This exercise showed that enforcing constraints can lead to diverse outcomes, depending on the specific data and model: improved predictions, no significant change, or slightly worse predictions.

We hope that the introduction of the EVI index will be a valuable metric to analyze density functionals, providing an insightful measure that may reveal hidden trends in DFT approximations.

\begin{acknowledgement}

This project has been supported by the Department of Energy through the grant DE-SC0022241.

\end{acknowledgement}

\bibliography{refs_1}

\end{document}